\documentclass[a4paper]{article}\usepackage{anysize,times}\usepackage{rsflash}

\usepackage{cite}\usepackage{epsfig}

\marginsize{1in}{1in}{1in}{1in}

\begin{document}\begin{titlepage}
\large\renewcommand{\thefootnote}{\normalsize\fnsymbol{footnote}}\hfill
\begin{tabular}{l}HEPHY-PUB 789/04\\%UWThPh-2004-??\\
July 2004\end{tabular}
\\[2cm]\Huge\begin{center}{\Huge\bf FACETS OF THE\\SPINLESS SALPETER
EQUATION}\\\vspace{2cm}\Large{\bf Wolfgang LUCHA\footnote[1]{\large\ {\em
E-mail address\/}: wolfgang.lucha@oeaw.ac.at}}\\[.3cm]\large Institut f\"ur
Hochenergiephysik,\\\"Osterreichische Akademie der Wissenschaften,\\
Nikolsdorfergasse 18, A-1050 Wien, Austria\\[1cm]\Large{\bf Franz
F.~SCH\"OBERL\footnote[2]{\large\ {\em E-mail address\/}:
franz.schoeberl@univie.ac.at}}\\[.3cm]\large Institut f\"ur Theoretische
Physik, Universit\"at Wien,\\Boltzmanngasse 5, A-1090 Wien, Austria

\vspace{1cm}{\large\bf Abstract}\end{center}\large The spinless Salpeter
equation represents the simplest, and most straightforward, generalization of
the Schr\"odinger equation of standard nonrelativistic quantum theory towards
the inclusion of relativistic kinematics. Moreover, it can be also regarded
as a well-defined approximation to the Bethe--Salpeter formalism for
descriptions of bound states in relativistic quantum field theories. The
corresponding Hamiltonian is, in contrast to all Schr\"odinger operators, a
nonlocal operator. Because of the nonlocality, constructing analytical
solutions for such kind of equation of motion proves difficult. In view of
this, different sophisticated techniques have been developed in order to
extract rigorous analytical information about these solutions. This review
introduces some of these methods and compares their significance by
application to interactions relevant in physics.

\vfill

\renewcommand{\thefootnote}{\arabic{footnote}}\normalsize\end{titlepage}

\twocolumn

\marginsize{1in}{1in}{1in}{1in}

\section{Bethe--Salpeter Formalism in the ``Instantaneous Approximation''}
Within quantum field theory, the appropriate framework for the description of
bound states is the Bethe--Salpeter formalism \cite{BSE}. Therein, all bound
states of two particles (in fact, of any two fermionic constituents) are
governed by the {\em homogeneous Bethe--Salpeter equation\/}. Here we are
interested in a particular well-defined approximation to this formalism,
obtained by several simplifying steps:\begin{enumerate}\item The {\em
instantaneous approximation\/}, neglecting any retardation effect, considers
all interactions of the (two) bound-state constituents in their static
limit.\item The additional assumption that all the bound-state constituents
propagate as free particles with some effective mass $m$ yields the {\em
Salpeter equation\/} \cite{SE}.\item A disregard of all of their spin degrees
of freedom focuses on the treatment of scalar bound particles.\item In
technical respect, the {\em canonical transformation\/}
\begin{equation}\mbox{\boldmath{$x$}}\to\lambda\,\mbox{\boldmath{$x$}}\
,\quad\mbox{\boldmath{$p$}}\to\frac{\mbox{\boldmath{$p$}}}{\lambda}
\label{Eq:scale}\end{equation}of position ($\mbox{\boldmath{$x$}}$) and
momentum ($\mbox{\boldmath{$p$}}$) variables casts in the case of particles
of equal mass $m$ for a scale factor $\lambda=2$ this approach into
one-particle form.\end{enumerate}(For more details of the derivation,
consult, for instance, Refs.~\cite{Lucha91:BSQ,Resag94,Kopaleishvili01} and
references therein.) Refraining from the nonrelativistic limit, we get the
(nonlocal!) Hamiltonian\begin{equation}H=T+V\ .\label{Eq:SRH}\end{equation}
This operator is composed of the ``square-root operator'' $T$ of the
relativistically correct expression for the kinetic or free energy of a
particle of mass $m$ and momentum~$\mbox{\boldmath{$p$}},$\begin{equation}
T=T(\mbox{\boldmath{$p$}})\equiv\sqrt{\mbox{\boldmath{$p$}}^2+m^2}\
,\label{Eq:RKE}\end{equation}and a (coordinate-dependent) static interaction
potential$$V=V(\mbox{\boldmath{$x$}})\ ;$$frequently, the potential
$V(\mbox{\boldmath{$x$}})$ is assumed to be a central potential that depends
merely on the radial coordinate $r$:$$V=V(r)\ ,\quad
r\equiv|\mbox{\boldmath{$x$}}|\ .$$The eigenvalue equation for this
particular Hamiltonian,\begin{equation}H\,|\chi_k\rangle=E_k\,|\chi_k\rangle\
,\quad k=0,1,2,\dots\ ,\label{Eq:EVE}\end{equation}defining a complete system
of Hilbert-space eigenstates $|\chi_k\rangle$ of $H$ corresponding to its
(energy) eigenvalues
$E_k,$$$E_k\equiv\frac{\langle\chi_k|\,H\,|\chi_k\rangle}
{\langle\chi_k|\chi_k\rangle}\ ,\quad k=0,1,2,\dots\ ,$$is commonly known as
the ``spinless Salpeter equation.''

\section{The Relativistic Virial Theorem}\label{Sec:RVT}Useful general
statements about the solutions of explicit or implicit eigenvalue equations
may be proved with the help of virial theorems obtained by generalization
\cite{Lucha:RVT} of the well-known result of nonrelativistic quantum theory.
(Ref.~\cite{Lucha:RVTs} is a brief review of relativistic virial theorems.)
For eigenvalue equations of the form (\ref{Eq:EVE}), the derivations of such
virial theorems can be traced back to the (trivial) observation that
expectation values taken with respect to given eigenstates $|\chi_k\rangle$
of $H$ --- or matrix elements taken with respect to arbitrary pairs of {\em
degenerate\/} eigenstates, $|\chi_i\rangle$ and $|\chi_j\rangle,$ of $H,$
i.~e., eigenstates satisfying $E_i=E_j$ --- of the commutators $[G,H]$ of the
operator $H$ and any other operator $G$ (the domain of which must be assumed
to contain the domain of $H$) clearly vanish. Suppressing the subscript that
enumerates the eigenstates, this
means\begin{equation}\langle\chi|\,[G,H]\,|\chi\rangle=0\ .\label{Eq:EV(C)}
\end{equation}For the symmetrized, self-adjoint generator of dilations,
\begin{equation}G\equiv\mbox{$\frac{1}{2}$}\,(\mbox{\boldmath{$x$}}\cdot
\mbox{\boldmath{$p$}}+\mbox{\boldmath{$p$}}\cdot\mbox{\boldmath{$x$}})\
,\label{Eq:DilGen}\end{equation}and $H$ of the form (\ref{Eq:SRH}) their
commutator $[G,H]$ becomes$$[G,H]={\rm
i}\left[\mbox{\boldmath{$p$}}\cdot\frac{\partial\,T}
{\partial\mbox{\boldmath{$p$}}}(\mbox{\boldmath{$p$}})-\mbox{\boldmath{$x$}}
\cdot\frac{\partial\,V}{\partial\mbox{\boldmath{$x$}}}(\mbox{\boldmath{$x$}})
\right].$$In this case Eq.~(\ref{Eq:EV(C)}) yields the {\em master virial
theorem\/} \cite{Lucha:RVT,Lucha:RVTs}
\begin{equation}\left\langle\chi\left|\,\mbox{\boldmath{$p$}}\cdot
\frac{\partial\,T}{\partial\mbox{\boldmath{$p$}}}(\mbox{\boldmath{$p$}})\,
\right|\chi\right\rangle=\left\langle\chi\left|\,\mbox{\boldmath{$x$}}\cdot
\frac{\partial\,V}{\partial\mbox{\boldmath{$x$}}}(\mbox{\boldmath{$x$}})\,
\right|\chi\right\rangle;\label{Eq:RVT}\end{equation}this relation expresses
the equality of all the expectation values of the momentum-space radial
derivative of $T(\mbox{\boldmath{$p$}})$ and the (configuration-space) radial
derivative of $V(\mbox{\boldmath{$x$}}),$ it produces the specific virial
theorem for a particular~$H.$ For any nonrelativistic (Schr\"odinger)
Hamiltonian, i.~e.,$$H=H_{\rm
S}=m+\frac{\mbox{\boldmath{$p$}}^2}{2\,m}+V(\mbox{\boldmath{$x$}})\ ,$$
Theorem (\ref{Eq:RVT}) entails, retaining the conventional
factor~$\mbox{$\frac{1}{2}$},$
\begin{equation}\left\langle\chi\left|\,\frac{\mbox{\boldmath{$p$}}^2}{2\,m}\,
\right|\chi\right\rangle=\frac{1}{2}\left\langle\chi\left|\,
\mbox{\boldmath{$x$}}\cdot\frac{\partial\,V}{\partial\mbox{\boldmath{$x$}}}
(\mbox{\boldmath{$x$}})\,\right|\chi\right\rangle.\label{Eq:NRVT}\end{equation}
For the semirelativistic ``spinless-Salpeter'' Hamiltonian (\ref{Eq:SRH}),
involving the square-root operator of the relativistic kinetic energy
(\ref{Eq:RKE}), our master virial theorem (\ref{Eq:RVT}) leads~to
\begin{equation}\left\langle\chi\left|\,\frac{\mbox{\boldmath{$p$}}^2}
{\sqrt{\mbox{\boldmath{$p$}}^2+m^2}}\,\right|\chi\right\rangle=
\left\langle\chi\left|\,\mbox{\boldmath{$x$}}\cdot\frac{\partial\,V}{\partial
\mbox{\boldmath{$x$}}}(\mbox{\boldmath{$x$}})\,\right|\chi\right\rangle.
\label{Eq:SVT}\end{equation}In the nonrelativistic limit $m\to\infty$ (i.~e.,
for $\mbox{\boldmath{$p$}}^2\ll m^2$), this {\em spinless-Salpeter
relativistic virial theorem\/}, Eq.~(\ref{Eq:SVT}), necessarily reduces to
its nonrelativistic counterpart (\ref{Eq:NRVT}). Similarly, the virial
theorem for the Dirac equation \cite{Fock30,Brack83} is easily deduced
\cite{Lucha:RVTs} from our master virial theorem (\ref{Eq:RVT}).

\section{Bounds to Energy Eigenvalues of a Spinless-Salpeter Hamiltonian}The
precise determination of eigenvalues of operators is of particular importance
for any formulation of quantum theory. Unfortunately, for most cases it is
not possible to determine the point spectrum (the set of all eigenvalues) of
a given operator {\em analytically\/}. Several powerful tools, however, allow
to derive analytic {\em bounds\/} to eigenvalues; applications of these
techniques to the spinless-Salpeter operator (\ref{Eq:SRH}) are reviewed, for
instance, in
Refs.~\cite{Lucha94:Como,Lucha98:Dubna,Lucha:Oberwoelz,Lucha:Dubrovnik}.

\subsection{Minimum--maximum principle}\label{Sec:MMP}The theoretical
foundation of any derivation of a system of rigorous upper bounds to the
(isolated) eigenvalues of some operator $H$ in Hilbert space and hence the
primary tool for any localization of the discrete spectrum of $H$ is the
well-known {\em minimum--maximum principle\/}
\cite{Reed78,Weinstein72,Thirring90}. Its precise formulation is based on
several prerequisites:\begin{itemize}\item Let this operator $H$ be some {\em
self-adjoint\/} operator.\item Assume that this operator is {\em bounded from
below\/}.\item Define the {\em eigenvalues\/} of $H,$ $E_k,$ $k=0,1,2,\dots,$
by the eigenvalue equation, with eigenstates $|\chi_k\rangle,$
$$H\,|\chi_k\rangle=E_k\,|\chi_k\rangle\ ,\quad k=0,1,2,\dots\ .$$\item Let
these eigenvalues $E_k$ be {\em ordered}, according~to$$E_0\le E_1\le
E_2\le\cdots\ .$$\item Consider only the eigenvalues $E_k$ {\em below\/} the
onset of the {\em essential spectrum\/} of the above operator $H.$\item
Restrict all considerations to some $d$-dimensional {\em subspace\/}
$D_d\subset{\cal D}(H)$ of the domain ${\cal D}(H)$ of~$H.$\end{itemize}Then
this theorem asserts that every eigenvalue $E_k$ of $H$ --- counting
multiplicity of degenerate levels ---
satisfies$$E_k\le\displaystyle\sup_{|\psi\rangle\in D_{k+1}}
\frac{\langle\psi|\,H\,|\psi\rangle}{\langle\psi|\psi\rangle}\quad\mbox{for
all}\ k=0,1,2,\dots\ .$$In the case of one-dimensional subspaces, that is,
$d=1,$ the minimum--maximum theorem reduces to {\em Rayleigh's principle\/}:
the ground-state eigenvalue $E_0$ of an operator $H$ is less than, or equal
to, every expectation value of
$H$:$$E_0\le\frac{\langle\psi|\,H\,|\psi\rangle}{\langle\psi|\psi\rangle}\
,\quad|\psi\rangle\in{\cal D}(H)\ .$$Given some operator inequality satisfied
by the operator $H,$ the minimum--maximum principle may be employed to
derive, by comparison, upper bounds on the (discrete) eigenvalues of $H,$
provided that a few assumptions hold:\begin{itemize}\item The operator $H,$
exhibiting all properties required by the minimum--maximum principle, is
bounded from above by some other operator called ${\cal O},$ i.~e., it is
subject to an {\em (operator) inequality\/} of the form$$H\le{\cal O}\
.$$Applying both the minimum--maximum principle and this operator inequality,
any eigenvalue $E_k$ of $H$ must be bounded from above by the supremum of the
expectation values of the operator ${\cal O}$ within the $(k+1)$-dimensional
subspace $D_{k+1}$ of ${\cal
D}(H)$:\begin{eqnarray}E_k&\equiv&\frac{\langle\chi_k|\,H\,|\chi_k\rangle}
{\langle\chi_k|\chi_k\rangle}\nonumber\\&\le&\sup_{|\psi\rangle\in D_{k+1}}
\frac{\langle\psi|\,H\,|\psi\rangle}{\langle\psi|\psi\rangle}\nonumber\\&\le&
\sup_{|\psi\rangle\in D_{k+1}}\frac{\langle\psi|\,{\cal O}\,|\psi\rangle}
{\langle\psi|\psi\rangle}\nonumber\\&&\quad\mbox{for all}\ k=0,1,2,\dots\
.\label{Eq:EVI}\end{eqnarray}\item All eigenvalues $\widehat E_k$ of ${\cal
O}$ are {\em ordered\/} according to$$\widehat E_0\le\widehat E_1\le\widehat
E_2\le\cdots\ .$$\item Every $k$-dimensional subspace $D_k$ in the chain of
inequalities which constitutes Eq.~(\ref{Eq:EVI}) is spanned by the first $k$
eigenvectors of the operator ${\cal O},$ or by precisely those eigenvectors
of ${\cal O}$ that correspond to the first $k$ eigenvalues $\widehat
E_0,\widehat E_1,\dots,\widehat E_{k-1}$ of our ${\cal O}.$ For this case it
is very easy to convince oneself that the supremum of all expectation values
of the operator ${\cal O}$ over the $(k+1)$-dimensional subspace $D_{k+1}$ is
then identical to the eigenvalue $\widehat E_k$ of ${\cal O}$:
$$\sup_{|\psi\rangle\in D_{k+1}}\frac{\langle\psi|\,{\cal O}\,|\psi\rangle}
{\langle\psi|\psi\rangle}=\widehat E_k\ .$$\end{itemize}Consequently, an
eigenvalue $E_k,$ $k=0,1,2,\dots,$ of the discrete spectrum of $H(\le{\cal
O})$ is bounded from above~by the corresponding eigenvalue $\widehat E_k,$
$k=0,1,2,\dots,$ of ${\cal O}$:$$E_k\le\widehat E_k\quad\mbox{for all}\
k=0,1,2,\dots\ .$$It remains to prove an ``appropriate'' operator inequality.
(Summaries of the idea to find bounds by combining the minimum--maximum
principle with reasonable operator inequalities may be found, e.~g., in
Refs.~\cite{Lucha96rcpaubel,Lucha:Oberwoelz,Lucha:Dubrovnik,Lucha99-1dimsrcp}.)

\subsection{Analytical upper bounds}

\subsubsection{The trivial nonrelativistic Schr\"odinger
bound}\label{Sec:NUB}The inequality $(T-m)^2\ge0$ expressing nothing but~the
{\em positivity\/} of the square of the operator $T-m$ may be, for $m>0,$
written as an inequality for the kinetic energy~$T$:$$T\le
m+\frac{\mbox{\boldmath{$p$}}^2}{2\,m}\ .$$(The right-hand side is the
tangent line to the square root in the relativistic kinetic energy $T$ in the
point of contact $\mbox{\boldmath{$p$}}^2=0.$) This result proves
\cite{Lucha96rcpaubel} that $H$ is bounded from above by a nonrelativistic
Schr\"odinger Hamiltonian $H_{\rm S}$:$$H\le H_{\rm
S}=m+\frac{\mbox{\boldmath{$p$}}^2}{2\,m}+V\ .$$For a pure Coulomb potential
$V(r)=-\alpha/r,$ the energy eigenvalues of the Schr\"odinger Hamiltonian
$H_{\rm S}$ depend only on the principal quantum number $n,$ related to both
radial and orbital angular-momentum quantum numbers by $n=n_{\rm r}+\ell+1,$
with $n_{\rm r}=0,1,2,\dots,$ $\ell=0,1,2,\dots$:$$E_{{\rm
S},n}=m\left(1-\frac{\alpha^2}{2\,n^2}\right).$$

\subsubsection{A ``squared'' bound}\label{Sec:QUB}A relation between the
(semirelativistic) Hamiltonian $H$ and a nonrelativistic Schr\"odinger
operator may be found \cite{Lucha96rcpaubel} by considering the square $H^2$
of $H$ and by realizing that the anticommutator $T\,V+V\,T$ of relativistic
kinetic energy $T$ and potential $V$ generated by the square
fulfils$$T\,V+V\,T\le\mbox{\boldmath{$p$}}^2+V^2+2\,m\,V\ ,$$as may be shown
\cite{Lucha96rcpaubel} by inspecting some consequences of the positivity of
the square of the operator
$T-m-V$:\begin{eqnarray*}H^2&=&T^2+V^2+T\,V+V\,T\\&\le&Q\equiv
2\,\mbox{\boldmath{$p$}}^2+m^2+2\,V^2+2\,m\,V\ .\end{eqnarray*}With this
inequality, the minimum--maximum principle, recalled in
Subsect.~\ref{Sec:MMP}, immediately guarantees that the energy eigenvalues
$E_k$ of $H$ are bounded from above by the square root of the corresponding
eigenvalues ${\cal E}_{Q,k}$~of the Schr\"odinger operator $Q,$ constructed
by squaring $H$:$$E_k\le\sqrt{{\cal E}_{Q,k}}\ ,\quad k=0,1,2,\dots\ .$$For
the case of a pure Coulomb potential $V(r)=-\alpha/r,$ the operator $Q$ has
the same structure as the Schr\"odinger Hamiltonian $H_{\rm S}$ of
Subsect.~\ref{Sec:NUB}, with $\ell$ replaced by an effective orbital angular
momentum quantum number $L$ involving both the usual $\ell$ and the Coulomb
coupling, $\alpha$:\begin{equation}L\,(L+1)=\ell\,(\ell+1)+\alpha^2\
,\quad\ell=0,1,2,\dots\ .\label{Eq:Leff}\end{equation}The set of eigenvalues
${\cal E}_Q$ of a ``Coulombic'' operator~$Q,$$${\cal
E}_{Q,N}=m^2\left(1-\frac{\alpha^2}{2\,N^2}\right),$$is determined by the
{\em effective\/} principal quantum number$$N=n_{\rm r}+L+1\ ,\quad n_{\rm
r}=0,1,2,\dots\ .$$Unfortunately, in the Coulomb case the squared bounds are
above, and thus worse than, the Schr\"odinger bounds.

\subsection{Rigorous semianalytical upper bound}\label{Sec:GUB}We regard an
energy bound as {\em semianalytical\/} if it can be derived by an (in
general, numerical) optimization of an analytically given expression over a
single real variable. Taking advantage, as a straightforward generalization
of the (simple) line of argument sketched in Subsect.~\ref{Sec:NUB}, of the
inequality $(T-\mu)^2\ge0$ requiring an arbitrary real parameter $\mu$ of
mass dimension 1 and obviously holding for all self-adjoint $T$
\cite{Martin88} implies, for the kinetic
energy,$$T\le\frac{\mbox{\boldmath{$p$}}^2+m^2+\mu^2}{2\,\mu}\quad\mbox{for
all}\ \mu>0\ .$$This translates \cite{Lucha96rcpaubel} to a set of
inequalities for $H,$ each of these involving a Schr\"odinger-like
Hamiltonian,~$\widehat H_{\rm S}(\mu)$:$$H\le\widehat H_{\rm
S}(\mu)=\frac{\mbox{\boldmath{$p$}}^2+m^2+\mu^2}{2\,\mu}+V\quad\mbox{for
all}\ \mu>0\ .$$The best ``Schr\"odinger-like'' upper bound on any energy
eigenvalue $E_k$ of $H$ is then provided by the minimum of the
$\mu$-dependent energy eigenvalues of $\widehat H_{\rm S}(\mu),$ $\widehat
E_{{\rm S},k}(\mu)$:$$E_k\le\displaystyle\min_{\mu>0}\widehat E_{{\rm
S},k}(\mu)\ ,\quad k=0,1,2,\dots\ .$$For a pure Coulomb potential
$V(r)=-\alpha/r,$ the energy eigenvalues $\widehat E_{{\rm S},n}(\mu)$ of
$\widehat H_{\rm S}(\mu)$ read, with $n=n_{\rm r}+\ell+1,$$$\widehat E_{{\rm
S},n}(\mu)=\frac{1}{2\,\mu}\left[m^2+\mu^2\left(1-\frac{\alpha^2}{n^2}\right)
\right].$$Here, minimizing $\widehat E_{{\rm S},n}(\mu)$ with respect to
$\mu$ entails \cite{Lucha96rcpaubel}$$\displaystyle\min_{\mu>0}\widehat
E_{{\rm S},n}(\mu)=m\,\sqrt{1-\frac{\alpha^2}{n^2}}\ .$$This (exact) {\em
upper\/} bound \cite{Lucha96rcpaubel} to the energy eigenvalues of the
so-called ``spinless relativistic Coulomb problem'' holds for all those
values of the Coulomb coupling $\alpha$ for which the Hamiltonian $H$ with a
Coulomb potential can be regarded as a reasonable operator and arbitrary
levels of excitation, and for any value of the principal quantum number $n$
it definitely improves the Schr\"odinger bound:$$\min_{\mu>0}\widehat E_{{\rm
S},n}(\mu)<E_{{\rm S},n}\quad\mbox{for}\ \alpha\neq 0\ .$$Clearly, fixing
$\mu=m$ recovers the Schr\"odinger bounds.

\subsection{Exact semianalytical upper and lower bounds from the ``envelope
technique''}\label{Sec:EnvULB}Rigorous semianalytical expressions for both
upper and lower bounds to the eigenvalues $E_{n\ell}$ of the Hamiltonian $H$
are found by a geometrical operator comparison in an approach called
``envelope theory.'' The envelope theory constructs bounds on $E_{n\ell}$ by
comparing the spectrum of $H$ with the one of a conveniently formulated
``tangential Hamiltonian'' $\widetilde H$ involving some ``basis potential''
$h(r),$$$\widetilde H=\sqrt{m^2+\mbox{\boldmath{$p$}}^2}+c\,h(r)
+\mbox{const.}\ ,\quad c>0\ ,$$for which sufficient spectral information
(i.~e., either the exact eigenvalues or suitable bounds on these) is known.
Let $V(r)$ be a smooth transformation $V=g(h)$ of $h(r),$ with definite
convexity of $g(h).$ After optimization with respect to the point of contact
of $V(r)$ and the tangential potential, this technique produces bounds on
$E_{n\ell}$:~lower bounds for $g(h)$ convex $(g''>0),$ and upper bounds for
$g(h)$ concave $(g''<0).$ Suppressing for the moment the quantum numbers
$n\ell,$ all these bounds on $E$ may be cast into a common generic form
\cite{Lucha00-HO,Lucha01-DMAIa,Lucha01-DMAIb,Lucha02-DMAII,Lucha02-sum,
Lucha02-CP} with the individual bounds discriminated by a dimensionless
parameter, $P$:\begin{equation}E\approx\min_{r>0}
\left[\sqrt{m^2+\frac{1}{r^2}}+V(P\,r)\right].\label{Eq:EnvBd}\end{equation}
Here, that cryptic sign of approximate equality indicates that for any
definite convexity of $g(h)$ all expressions on the right-hand side represent
a lower bound for a convex $g(h)$ and an upper bound for a concave $g(h).$
The value of the parameter $P$ used in Eq.~(\ref{Eq:EnvBd}) is determined
by~the algebraic structure of the interaction potential $V(r),$ and by its
convexity with respect to the basis potential, $h(r)$:\begin{itemize}\item
The {\em spinless relativistic Coulomb problem\/} posed by $V(r)=-\alpha/r$
is well-defined if its coupling $\alpha$ is constrained to $\alpha<2/\pi$
\cite{Herbst}. The bottom of the corresponding spectrum of $H$ (or, its
ground-state energy eigenvalue, $E_0$) is bounded from below by$$E_0\ge
m\,\sqrt{1-\frac{\alpha^2}{P^2}}\ ,$$with the lower-bound parameter $P$ given
either~by $P=2/\pi$ for $\alpha$ fulfilling $0\le\alpha<2/\pi$ \cite{Herbst},
or~by$$P\equiv
P(\alpha)=\sqrt{\mbox{$\frac{1}{2}$}\left(1+\sqrt{1-4\,\alpha^2}\right)}$$
for $0\le\alpha\le\mbox{$\frac{1}{2}$},$ which obviously covers the range
$$P(\mbox{$\frac{1}{2}$})=\mbox{$\frac{1}{\sqrt{2}}$}\le P(\alpha)\le P(0)=1\
,$$as derived by weakening
\cite{Lucha01-DMAIa,Lucha01-DMAIb,Lucha02-DMAII,Lucha02-CP} an improved lower
bound to $E_0$ valid only for $0\le\alpha\le\frac{1}{2}$ \cite{Martin89}.

\newpage

If $V(r)$ is a convex transform $V=g(h),$ $g''>0,$ of the Coulomb potential
$h(r)=-1/r,$ the above envelope approximation generates a {\em lower\/} bound
\cite{Lucha01-DMAIa,Lucha01-DMAIb,Lucha02-DMAII,Lucha02-CP} on the
ground-state eigenvalue $E_0$ (on the entire spectrum) of the Hamiltonian $H$
for any choice of the Coulomb lower-bound parameter $P.$

Clearly, the quoted upper bounds on the Coulomb coupling $\alpha$ apply also
to any ``effective'' Coulomb coupling in $\widetilde H.$ Consequently, they
translate into a constraint on all coupling constants introduced by the
interaction potential $V(r)$ under investigation. (An example for these
restrictions enforced by the Coulomb menace will be given in
Subsect.~\ref{Sec:FP}.)\item For $V(r)$ a concave transform $V=g(h),$
$g''<0,$ of the harmonic-oscillator potential $h(r)=r^2,$ a straightforward
application of the above envelope approximation yields {\em upper\/} bounds
\cite{Lucha01-DMAIa,Lucha01-DMAIb,Lucha02-DMAII,Lucha02-CP} to {\em all\/}
the eigenvalues $E_{n\ell}$ of the Hamiltonian $H;$ the parameter $P$ for a
given energy level identified by quantum numbers $n\ell$ is, in this case,
related to the explicitly algebraically known eigenvalues ${\cal E}_{n\ell}$
of the (nonrelativistic) Schr\"odinger operator
$\mbox{\boldmath{$p$}}^2+r^2$:$$P\equiv
P_{n\ell}(2)=\mbox{$\frac{1}{2}$}\,{\cal
E}_{n\ell}=2\,n+\ell-\mbox{$\frac{1}{2}$}\ .$$\item For $V(r)$ a concave
transform $V=g(h),$ $g''<0,$ of the linear potential $h(r)=r,$ the
application of a ``generalized'' envelope approximation provides {\em
upper\/} bounds \cite{Lucha02-DMAII,Lucha02-CP} to {\em all\/} eigenvalues
$E_{n\ell}$ of $H$ if the parameters $P$ which characterize the energy levels
are given, in terms of the eigenvalues ${\cal E}_{n\ell}$ of the
nonrelativistic Schr\"odinger operator $\mbox{\boldmath{$p$}}^2+r,$~by
\begin{equation}P\equiv P_{n\ell}(1)=2\left(\mbox{$\frac{1}{3}$}\,{\cal
E}_{n\ell}\right)^{3/2}\ ;\label{Eq:P(1)}\end{equation}the parameter values
$P_{n\ell}(1)$ corresponding to the lowest-lying energy levels can be found
in Table~\ref{Tab:P(1)} (for more details see, for instance,
Refs.~\cite{Lucha00-HO,Lucha01-DMAIa,Lucha01-DMAIb,Lucha02-DMAII}).
\end{itemize}\begin{table}[ht]\caption{Numerical values of the parameter
$P_{n\ell}(1)$ used in the linear-potential-based lower envelope bounds and
defined in Eq.~(\ref{Eq:P(1)}) for the lowest-lying energy levels
$n\ell.$}\label{Tab:P(1)}\vspace{1ex}
\begin{center}\begin{tabular}{ccr}\hline\hline&&\\[-1.5ex]
\multicolumn{1}{c}{$n$}&\multicolumn{1}{c}{$\ell$}&
\multicolumn{1}{c}{$P_{n\ell}(1)$}\\[1ex]\hline\\[-1.5ex]
1&0&1.37608\\2&0&3.18131\\3&0&4.99255\\4&0&6.80514\\5&0&8.61823\\[.5ex]
1&1&2.37192\\2&1&4.15501\\3&1&5.95300\\4&1&7.75701\\5&1&9.56408\\[.5ex]
1&2&3.37018\\2&2&5.14135\\3&2&6.92911\\4&2&8.72515\\5&2&10.52596\\[1ex]
\hline\hline\end{tabular}$\qquad$\begin{tabular}{ccr}\hline\hline&&\\[-1.5ex]
\multicolumn{1}{c}{$n$}&\multicolumn{1}{c}{$\ell$}&
\multicolumn{1}{c}{$P_{n\ell}(1)$}\\[1ex]\hline\\[-1.5ex]
1&3&4.36923\\2&3&6.13298\\3&3&7.91304\\4&3&9.70236\\5&3&11.49748\\[.5ex]
1&4&5.36863\\2&4&7.12732\\3&4&8.90148\\4&4&10.68521\\5&4&12.47532\\[.5ex]
1&5&6.36822\\2&5&8.12324\\3&5&9.89276\\4&5&11.67183\\5&5&13.45756\\[1ex]
\hline\hline\end{tabular}\end{center}\end{table}If the potential $V(r)$ is
the sum of several distinct terms,$$V(r)=\sum_iV_i(r)\ ,\quad
V_i(r)=c_i\,h_i(r)\ ,$$where every {\em component problem\/} defined by the
operator$$\sqrt{m^2+\mbox{\boldmath{$p$}}^2}+c_i\,h_i(r)$$supports, for a
sufficiently large $c_i,$ a discrete eigenvalue $E_{i,0}$ at the bottom of
its spectrum and information about the lowest energy eigenvalue, $E_{i,0},$
is available, all these pieces of information can be combined to a lower
bound to $E_0$ \cite{Lucha02-sum}; for sums of pure power-law terms ${\rm
sgn}(q)\,r^q,$\begin{equation}V_{\rm PL}(r)=\sum_qa(q)\,{\rm sgn}(q)\,r^q\
,\label{Eq:PLP}\end{equation}where the coefficients $a(q)$ of the pure
power-law terms, ${\rm sgn}(q)\,r^q,$ in the potential are positive, that is,
$a(q)\ge0,$ and do not vanish {\em all\/}, this yields the ``sum lower
bound''$$E_0\ge\min_{r>0}\left[\sqrt{m^2+\frac{1}{r^2}}+\sum_qa(q)\,{\rm
sgn}(q)\,(\underline{P}(q)\,r)^q\right]$${\em provided\/} that some set of
lower-bound parameters $\underline{P}(q)$ can be derived such that, whenever
$V(r)$ consists of just one single component, the above inequality yields
either the corresponding exact ground-state energy eigenvalue or, at least, a
rigorous lower bound to this latter quantity:\begin{itemize}\item For Coulomb
components, that is, $h_i(r)=-1/r,$ $\underline{P}(-1)$ is the Coulomb
lower-bound parameter $P.$\item For linear components, that is, $h_i(r)=r,$
$\underline{P}(1)$ is derived from the lowest eigenvalue ${\cal E}_0$ of
$\sqrt{\mbox{\boldmath{$p$}}^2}+r,$$$\underline{P}(1)=\mbox{$\frac{1}{4}$}\,
{\cal E}_0^2=1.2457\ .$$\end{itemize}It is straightforward to (try to)
generalize these envelope techniques from the simpler one-body case
summarized in this review to systems composed of arbitrary numbers of
relativistically moving interacting particles described by a semirelativistic
spinless Salpeter equation \cite{Lucha01-NHO,Lucha03-NHO-m=0,Lucha04-NV(r2)}.
At least for the particular case of all \mbox{harmonic-oscillator} potentials
$V(r)=c\,r^2$ with $c>0$ the generalized upper bounds presented in
Subsect.~\ref{Sec:GUB} and the envelope upper bounds can be shown to be
equivalent to each other \cite{Lucha02-DMAII}.

\subsection{Rayleigh--Ritz (variational) technique}\label{Sec:RRVT}An
immediate consequence of the minimum--maximum principle is the
``Rayleigh--Ritz (variational) technique:''\begin{itemize}\item Introduce the
{\em restriction\/} $\widehat H$ of some operator $H$ to a subspace $D_d$ by
orthogonal projection $P$ to~$D_d$:$$\left.\widehat H\equiv
H\right|_{D_d}:=P\,H\,P\ .$$\item Identify all $d$ {\em eigenvalues\/}
$\widehat E_k,$ $k=0,1,\dots,d-1,$ of the restricted operator $\widehat H$ as
the solutions of the eigenvalue equation of $\widehat H$ for the eigenstates
$|\widehat\chi_k\rangle$:$$\widehat H\,|\widehat\chi_k\rangle=\widehat
E_k\,|\widehat\chi_k\rangle\ ,\quad k=0,1,\dots,d-1\ .$$\item Let these
eigenvalues $\widehat E_k$ be {\em ordered}, according~to$$\widehat
E_0\le\widehat E_1\le\cdots\le\widehat E_{d-1}\ .$$\end{itemize}Then every
(discrete) eigenvalue $E_k$ of $H$ --- if counting the multiplicity of
degenerate levels --- is bounded from above by the eigenvalue $\widehat E_k$
of the restricted operator~$\widehat H$:$$E_k\le\widehat E_k\quad\mbox{for
all}\ k=0,1,\dots,d-1\ .$$If that $d$-dimensional subspace $D_d$ is spanned
by any set of $d$ (of course, linearly independent) basis vectors
$|\psi_k\rangle,$ $k=0,1,\dots,d-1,$ the eigenvalues $\widehat E_k$ can
immediately be determined, by the diagonalization of the $d\times d$
matrix$$\left(\langle\psi_i|\,\widehat H\,|\psi_j\rangle\right),\quad
i,j=0,1,\dots,d-1\ ,$$that is, as the $d$ roots of the characteristic
equation of~$\widehat H,$\begin{eqnarray*}\det\left(\langle\psi_i|\,\widehat
H\,|\psi_j\rangle-\widehat E\,\langle\psi_i|\psi_j\rangle\right)=0\ ,\\
i,j=0,1,\dots,d-1\ .\,\end{eqnarray*}To establish this, expand any
eigenvector $|\widehat\chi_k\rangle$ of $\widehat H$ over the basis
$\{|\psi_i\rangle,\ i=0,1,\dots,d-1\}$ of the subspace $D_d.$

\subsection{Variational upper bounds}\label{Sec:VUB}The {\em quality\/}
achieved by the variational solution of some eigenvalue problem depends
decisively on the definition of the trial subspace $D_d$ employed by the
Rayleigh--Ritz technique briefly sketched in Subsect.~\ref{Sec:RRVT}:
enlarging $D_d$ to higher dimensions $d$ or choosing a more sophisticated
basis $\{|\psi_i\rangle,\ i=0,1,\dots,d-1\}$ which spans $D_d$ will, in
general, increase the {\em accuracy\/} of the obtained solutions.

For spherically symmetric (central) potentials $V(r),$ that is, for all
potentials which depend only on the radial coordinate
$r\equiv|\mbox{\boldmath{$x$}}|,$ a convenient and thus rather popular choice
for the basis vectors $\{|\psi_i\rangle,\ i=0,1,\dots,d-1\}$ is that one the
configuration-space representation of which involves the complete orthogonal
system of generalized Laguerre polynomials
\cite{Jacobs86,Lucha:LagB,Lucha:Oberwoelz,Lucha:Dubrovnik} ---
cf.~Appendix~\ref{App:Lag}.

\newpage\noindent In the one-dimensional case \cite{Lucha94:VA-SRCP} realized
in the notation of Appendix~\ref{App:Lag} if all quantum numbers
$k=\ell=m=0,$ the Laguerre basis collapses to just a single basis vector:
$$\psi(\mbox{\boldmath{$x$}})\equiv\psi_{0,00}(\mbox{\boldmath{$x$}})
=\sqrt{\frac{\mu^3}{\pi}}\,\exp(-\mu\,r)\ ,\quad\mu>0\ .$$With a trial state
$|\psi\rangle$ represented by this exponential and the trivial (nevertheless
fundamental) general inequality$$\frac{|\langle\psi|\,{\cal
O}\,|\psi\rangle|}{\langle\psi|\psi\rangle}\le\sqrt{\frac{\langle\psi|\,{\cal
O}^2\,|\psi\rangle}{\langle\psi|\psi\rangle}}\ ,$$which holds for any
self-adjoint, but otherwise arbitrary, operator ${\cal O}$ (${\cal
O}^\dagger={\cal O}$), Rayleigh's principle entails,~after optimization with
respect to the variational parameter $\mu,$ for a Coulomb potential
$V(r)=-\alpha/r$ the upper bound$$E_0\le m\,\sqrt{1-\alpha^2}\ ;$$this is
identical to the ``generalized'' upper energy bound on the ground-state or
$n=1$ eigenvalue of the Coulomb operator $H$ found by different reasoning in
Subsect.~\ref{Sec:GUB}.

\subsection{Application to illustrative interactions}Let us appreciate the
above bounds' beauty at examples.

\subsubsection{Trivial ``testing ground:'' Coulomb potential}\label{Sect:CP}
Our first example clearly must be the Coulomb
potential$$V(r)=-\frac{\alpha}{r}\ ,\quad\alpha>0\ ;$$this potential arises
from the exchange of some massless boson between the interacting objects.
Therefore it is of particular interest in many areas of physics. Its
effective interaction strength is given by a coupling $\alpha,$ identical~to
the fine structure constant in electrodynamics. We study\begin{itemize}\item
the somewhat naive nonrelativistic (Schr\"odinger) upper bound given in
Subsect.~\ref{Sec:NUB}, equivalent to a tangent line to the relativistic
kinetic operator $T,$\item the upper bound of Subsect.~\ref{Sec:QUB},
constructed by considering just the square of the Hamiltonian $H,$\item the
semianalytical upper bound of Subsect.~\ref{Sec:GUB}, as derived by
generalizing the idea of Subsect.~\ref{Sec:NUB},\item all three envelope
bounds of Subsect.~\ref{Sec:EnvULB}, namely,\begin{itemize}\item the
harmonic-oscillator-based upper bound,\item the upper bound involving a
linear potential,\item the lower bound obtained by ``loosening'' an absolute
lower bound on the spectrum of the ``semirelativistic Coulomb operator'' $H,$
and\end{itemize}\item the ``Rayleigh--Ritz'' upper bound of
Subsect.~\ref{Sec:VUB}.\end{itemize}

\newpage\noindent For the Coulomb potential under study, the optimization
required by the envelope bounds (\ref{Eq:EnvBd}) may be performed
analytically, yielding a result of precisely the same form as the generalized
upper bounds derived in Subsect.~\ref{Sec:GUB}, or as the squared upper
bounds proved in Subsect.~\ref{Sec:QUB}:\begin{equation}E_0(P)\approx
m\,\sqrt{1-\frac{\alpha^2}{P^2}}\ ,\label{Eq:CPB}\end{equation}where for the
ground state characterized by the quantum numbers $n=1,$ $\ell=0$ the
(single) parameter $P$ is given,\begin{itemize}\item for the ``Coulomb lower
bound'' (Subsect.~\ref{Sec:EnvULB}), by$$P\equiv P_{\rm
C}=P(\alpha)=\sqrt{\mbox{$\frac{1}{2}$}\left(1+\sqrt{1-4\,\alpha^2}\right)}\
,$$\item for the generalized upper bound
(Subsect.~\ref{Sec:GUB}),~by$$P\equiv P_{\rm G}=n=1\ ,$$\item for the
``linear upper bound'' (Subsect.~\ref{Sec:EnvULB}), as can be simply read off
from the first row in Table~\ref{Tab:P(1)},~by$$P\equiv P_{\rm
L}=P_{10}(1)=1.37608\ ,$$\item for the ``squared upper bound''
(Subsect.~\ref{Sec:QUB}), in accordance with the solution of
Eq.~(\ref{Eq:Leff}) for $L,$~by$$P\equiv P_{\rm
Q}=\sqrt{2}\,N=\frac{1+\sqrt{1+4\,\alpha^2}}{\sqrt{2}}\ ,$$\item and, in the
case of the ``harmonic-oscillator upper bound'' (Subsect.~\ref{Sec:EnvULB}),
from the $P_{n\ell}(2)$ results,~by$$P\equiv P_{\rm
H}=P_{10}(2)=\mbox{$\frac{3}{2}$}\ .$$\end{itemize}It is a very trivial
observation that, for fixed values of the Coulomb coupling, the ground-state
energy bounds (\ref{Eq:CPB}) are (monotone) increasing with increasing
parameter $P$:$$\frac{\partial\,E_0(P)}{\partial P}\ge0\ .$$Thus it is
straightforward to convince oneself that all the Coulomb ($E_{\rm C}$),
generalized ($E_{\rm G}$), nonrelativistic ($E_{\rm N}$), linear ($E_{\rm
L}$), squared ($E_{\rm Q}$) and harmonic-oscillator ($E_{\rm H}$) bounds on
the ground-state energy eigenvalue $E_0$ of the semirelativistic Coulomb
Hamiltonian $H$ have to satisfy\begin{eqnarray*}E_{\rm C}\le E_0\le E_{\rm
G}\le E_{\rm N}\le E_{\rm L}\le E_{\rm Q}\le E_{\rm H}\\\mbox{for}\
\alpha\le\alpha_0\equiv\sqrt{\mbox{$\frac{3}{8}$}\left(3-2\,\sqrt{2}\right)}\
,\\E_{\rm C}\le E_0\le E_{\rm G}\le E_{\rm N}\le E_{\rm L}\le E_{\rm H}\le
E_{\rm Q}\\\mbox{for}\
\alpha\ge\alpha_0\equiv\sqrt{\mbox{$\frac{3}{8}$}\left(3-2\,\sqrt{2}\right)}\
,\end{eqnarray*}taking into account the crossing of the upper bounds $E_{\rm
H}$ and $E_{\rm Q}$ at $\alpha_0^2=\frac{3}{8}\,(3-2\,\sqrt{2}),$ i.~e.,
$E_{\rm Q}(\alpha_0)=E_{\rm H}(\alpha_0).$

\newpage

For Coulomb-like interactions the only dimensional quantity among the
parameters of this theory is the mass $m$ of the interacting particles.
Consequently, in this case all energy eigenvalues are proportional to $m$:
the energy scale is set by $m.$ The ratio $E_k/m$ is a universal function of
the coupling $\alpha;$ w.~l.~o.~g.\ it thus suffices to fix $m=1.$

\begin{figure}[h]\begin{center}\psfig{figure=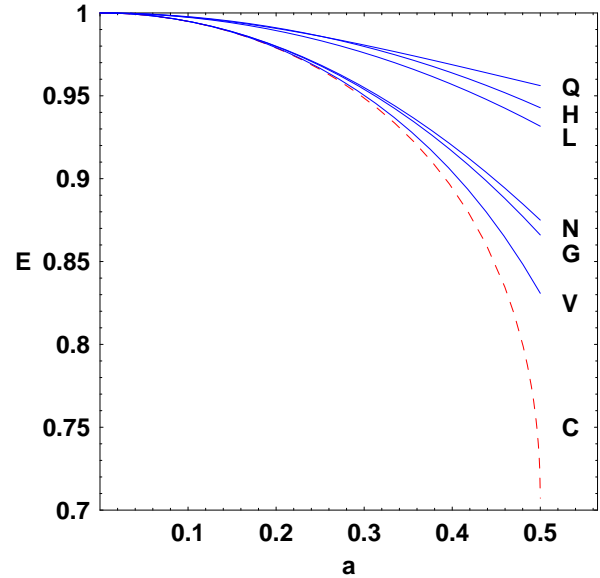,scale=0.76789}
\caption{Both upper ({\sl full lines\/}) and lower ({\sl dashed line\/})
bounds on the ground-state energy eigenvalue ({\bf E}) of the
semirelativistic Hamiltonian $H$ with Coulomb potential $V(r)=-a/r$ as a
function of the Coulomb coupling,~$a,$ for the s{\bf q}uared ({\bf Q}), {\bf
h}armonic-oscillator ({\bf H}), {\bf l}inear ({\bf L}), {\bf n}onrelativistic
({\bf N}), {\bf g}eneralized ({\bf G}), {\bf v}ariational ({\bf V}) and {\bf
C}oulomb ({\bf C}) [using the ``optimized'' $P(a)$]
approaches.}\label{Fig:CP}\end{center}\end{figure}

Figure~\ref{Fig:CP} compares for the ground state ($n_{\rm r}=\ell=0$) of the
spinless relativistic Coulomb problem the various bounds to the lowest energy
eigenvalue, $E_0,$ listed at the beginning of this subsection. Inspecting
Fig.~\ref{Fig:CP}, we note:\begin{itemize}\item the squared,
harmonic-oscillator, and linear upper bounds are numerically comparable to
each other;\item likewise the nonrelativistic and generalized upper bounds
are close to each other for all couplings~$\alpha;$\item using a Laguerre
trial space of dimension $d=25,$ the Rayleigh--Ritz variational upper bound
can be expected to come already pretty close to the exact eigenvalue $E_0$
--- which, in turn, clearly indicates that it is highly desirable to find
improvements for the lower bounds, in particular for large couplings $\alpha$
(this stimulated, e.~g., the analysis of Ref.~\cite{Lucha96:CCC}).
\end{itemize}

\subsubsection{Coulomb-plus-linear (or ``funnel'') potential}\label{Sec:FP}
Within the field of elementary particle physics, quantum chromodynamics (QCD)
is generally accepted to be that relativistic quantum field theory that
describes all strong interactions between quarks and gluons by assigning the
so-called ``colour'' degrees of freedom to these particles. In the
instantaneous approximation inherent to all of the QCD-inspired quark
potential models developed for the purely phenomenological description of
experimentally observed hadrons, as bound states of {\em constituent\/}
quarks, the strong forces are assumed to derive from an effective potential
generating the bound states (this description of hadrons within the framework
of quark potential models involving either nonrelativistic or relativistic
kinematics is reviewed, for instance, in
Refs.~\cite{Lucha91:BSQ,Lucha92:QAQBS}.) The prototype of all ``realistic,''
that is, phenomenologically acceptable (static) interquark potentials $V(r)$
consists of the sum of\begin{itemize}\item a Coulomb contribution generated
by a one-gluon exchange between quark bound-state constituents (dominating
the potential at short distances $r$) and\item a linear term including all
nonperturbative effects (that dominates the potential at large distances
$r$).\end{itemize}The resulting interaction potential $V(r)$ is characterized
by a ``funnel-type'' Coulomb-plus-linear form; therefore it is called the
Coulomb-plus-linear, or funnel, potential. Upon factorizing off a constant
$v,$ which spans the range $0<v\le1$ in order to parametrize an overall
interaction strength, we (prefer to) analyze this potential in the form
\begin{equation}V(r)=-\frac{c_1}{r}+c_2\,r=v\left(-\frac{a}{r}+b\,r\right).
\label{Eq:FP}\end{equation}Clearly, given the overall coupling strength $v,$
the actual shape of this potential is fixed by the {\em ratio\/} of the
positive parameters $a>0$ and $b>0;$ the coupling constants that enter, on
the one hand, in the general expression (\ref{Eq:PLP}) for sums of pure
power-law terms and, on the other hand, in our funnel potential (\ref{Eq:FP})
must be identified according~to\begin{eqnarray*}a(-1)&\equiv&c_1\equiv
a\,v>0\ ,\\a(1)&\equiv&c_2\equiv b\,v>0\ .\end{eqnarray*}In view of the lack
of fully analytical bounds we explore\begin{itemize}\item the three ``basic''
envelope bounds of Subsect.~\ref{Sec:EnvULB}, distinguished by the adopted
basis potential, viz.,\begin{itemize}\item the upper bound from a harmonic
oscillator,\item the upper bound involving a linear potential,\item the lower
bound due to a Coulomb potential,\end{itemize}\item the envelope {\em sum
lower bound\/}, derived in the sum approximation recalled by
Subsect.~\ref{Sec:EnvULB}, as well as\item the ``Rayleigh--Ritz'' upper bound
of Subsect.~\ref{Sec:VUB}.\end{itemize}

\newpage\noindent For definiteness, let us fix the potential parameters $a$
and $b$ to $a=0.2,$ $b=0.5.$ As done in the Coulomb-potential example (in
Subsect.~\ref{Sect:CP}) in order to take advantage of upper and lower bounds,
we investigate the ground-state energy $E_0.$ The basic envelope bounds are
computed by application of Eq.~(\ref{Eq:EnvBd}), for the appropriate
parameter~$P$:\begin{itemize}\item for the ``harmonic-oscillator upper
bound'' we use$$P\equiv P_{\rm H}=P_{10}(2)=\mbox{$\frac{3}{2}$}\ ;$$\item
for the ``linear upper bound'' we find from Table~\ref{Tab:P(1)}$$P\equiv
P_{\rm L}=P_{10}(1)=1.37608\ ;$$\item for the ``Coulomb lower bound,'' in
order to derive the maximum value $P$ consistent with $0<v\le1,$ we are
forced to evaluate that ``Coulomb coupling constant constraint'' mentioned in
Subsect.~\ref{Sec:EnvULB}, in its form \cite{Lucha01-DMAIb,{Lucha02-DMAII}}
fixed by our funnel potential (\ref{Eq:FP}),
$$c_1+\frac{P^4}{1-P^2}\frac{c_2}{m^2}\le P\,\sqrt{1-P^2}\ ,$$for the maximum
values of $c_1$ and $c_2,$ which gives$$P\equiv P_{\rm
C}=0.728112397\quad\mbox{for}\ m=1\ .$$\end{itemize}The ``sum lower bound''
is extracted from the expression given explicitly in
Subsect.~\ref{Sec:EnvULB} for power-law potentials by insertion of the
lower-bound parameters $\underline{P}(q=\pm1)$:\begin{itemize}\item the
Coulomb lower-bound parameter $P(\alpha)$ leads, for the relevant maximum
coupling $\alpha=a=0.2,$ in the Coulomb term of the sum approximation
to$$\underline{P}(-1)=P(\alpha)=P(a)=0.9789063\ ;$$\item the lower-bound
parameter required for any linear part of sum potentials is copied from
Subsect.~\ref{Sec:EnvULB},$$\underline{P}(1)=1.2457\ .$$\end{itemize}As
before, the Rayleigh--Ritz or variational upper bound is found in a trial
space of dimension $d=25$ spanned by the generalized Laguerre basis
(summarized in App.~\ref{App:Lag}).

\begin{figure}[ht]\begin{center}\psfig{figure=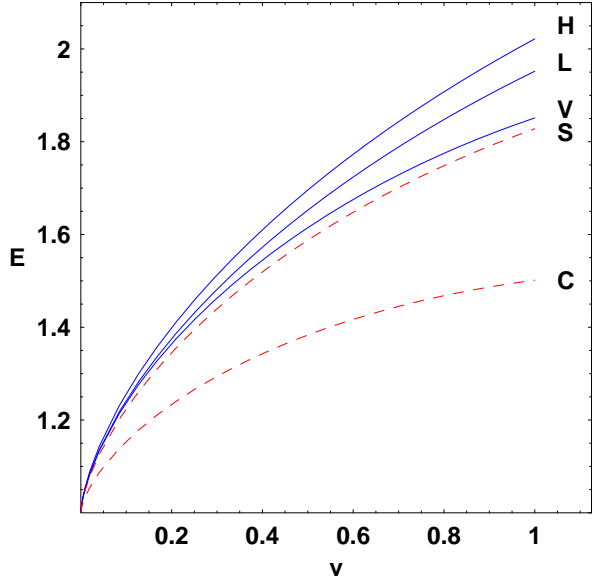,scale=0.76789}
\caption{Three upper ({\sl full lines\/}) and two lower ({\sl dashed
lines\/}) bounds on the ground-state energy eigenvalue ({\bf E}) of the
semirelativistic Hamiltonian $H$ with the so-called funnel potential
$V(r)=v\,(-a/r+b\,r),$ where $a=0.2,$ $b=0.5,$ $m=1.$ These include: the
harmonic-oscillator ({\bf H}), linear ({\bf L}) and variational ({\bf V})
upper bounds and the sum-approximation ({\bf S}) and Coulomb ({\bf C}) lower
bounds.}\label{Fig:FP}\end{center}\end{figure}

Figure~\ref{Fig:FP} depicts the bounds to the lowest eigenvalue $E_0$ of $H$
as function of the overall coupling strength $v$~in the funnel potential
(\ref{Eq:FP}). Remarkably, variational upper and sum lower bounds now
restrict $E_0$ to a narrow band.

\section{Approximate Solutions: Quality}Having determined --- for instance,
by application of the Rayleigh--Ritz technique sketched in
Subsect.~\ref{Sec:RRVT} --- for some $k=0,1,2,\dots$ the state
$|\widehat\chi_k\rangle\in D_d$ corresponding to any upper bound $\widehat
E_k$ on the exact eigenvalue $E_k$ of $H,$ one question immediately arises:
how closely resembles the approximate solution $|\widehat\chi_k\rangle$ the
exact eigenstate $|\chi_k\rangle?$

\newpage\noindent Standard criteria, such as the (relative) distance between
$\widehat E_k$ and true $E_k,$ or the overlap of approximate and exact
eigenstates, require the knowledge of the exact solution. In contrast to
this, the virial theorem (Sect.~\ref{Sec:RVT}) represents an indicator for
the accuracy of approximate eigenstates that merely uses information provided
by the variational approach: Since all eigenstates of $H$ satisfy any
relation of the form (\ref{Eq:EV(C)}), a significant imbalance in
Eq.~(\ref{Eq:RVT}) reveals that this approximation is far from optimum
\cite{Lucha:Q/A,Lucha:A/Q,Lucha02-DMAII}. Of course, because of the
involvement (\ref{Eq:EV(C)}) of the dilation generator (\ref{Eq:DilGen}) in
the derivation of Eq.~(\ref{Eq:RVT}), any variational solution found by
minimization of expectation values of $H$ with respect to the scale
transformations, or dilations, (\ref{Eq:scale}) will necessarily satisfy our
master virial theorem (\ref{Eq:RVT}).

\section{Summary, Concluding Remarks}The various efficient approaches
presented here allow to analyze the semirelativistic Hamiltonians of the
spinless Salpeter equation analytically; this is crucial for general
considerations that aim to answer questions of principle, like operator
boundedness. For numerical methods, see, for instance,
Refs.~\cite{Lucha92:MAP,Lucha:SAMM,{Lucha94:Como}} and the references
therein.

\appendix\section{The Generalized Laguerre Basis}\label{App:Lag}Assume every
basis function of $L_2(R^3)$ to factorize into a function of the radial
variable and the angular term. Its configuration-space representation has the
general form$$\psi_{k,\ell m}(\mbox{\boldmath{$x$}})=\Phi_{k,\ell}(r)\,{\cal
Y}_{\ell m}(\Omega_r)\ ,\quad r\equiv|\mbox{\boldmath{$x$}}|\ ;$$the
spherical harmonics ${\cal Y}_{\ell m}(\Omega)$ for angular momentum $\ell$
and projection $m$ depend on the solid angle $\Omega\equiv(\theta,\phi)$ and
satisfy a well-known orthonormalization condition:$$\int{\rm d}\Omega\,{\cal
Y}^\ast_{\ell m}(\Omega)\,{\cal
Y}_{\ell'm'}(\Omega)=\delta_{\ell\ell'}\,\delta_{mm'}\ .$$

The most popular choice
\cite{Jacobs86,Lucha:LagB,Lucha:Oberwoelz,Lucha:Dubrovnik} for the basis
states which span the Hilbert space $L_2(R^+)$ of [with the weight
$w(x)=x^2$] square-integrable functions $f(x)$ on the positive real line
$R^+$ --- which is the Hilbert space of radial trial functions
$\Phi_{k,\ell}(r)$ --- involves the generalized Laguerre polynomials
$L_k^{(\gamma)}(x),$ for parameter $\gamma$ \cite{Abramowitz,Bateman}:
$$\Phi_{k,\ell}(r)=N_{k,\ell}^{(\mu,\beta)}\,r^{\ell+\beta-1}\exp(-\mu\,r)\,
L_k^{(2\,\ell+2\,\beta)}(2\,\mu\,r)\ ;$$these generalized Laguerre
polynomials for parameter $\gamma$ are orthogonal polynomials, defined by the
power series$$L_k^{(\gamma)}(x)=\sum_{t=0}^k\,(-1)^t\left(\begin{array}{c}
k+\gamma\\k-t\end{array}\right)\frac{x^t}{t!}\ ,\quad k=0,1,\dots\ ,$$and
orthonormalized with weight function
$x^\gamma\exp(-x)$:\begin{eqnarray*}&&\int\limits_0^\infty{\rm
d}x\,x^\gamma\exp(-x)\,L_k^{(\gamma)}(x)\,L_{k'}^{(\gamma)}(x)\\
&&=\frac{\Gamma(\gamma+k+1)}{k!}\,\delta_{kk'}\ ,\quad k,k'=0,1,\dots\
.\end{eqnarray*}

The basis states defined by the generalized-Laguerre choice for the radial
basis functions $\Phi_{k,\ell}(r)$ involve two parameters, both of which may
be subsequently adopted for variational purposes: $\mu$ (with the dimension
of mass) and $\beta$ (dimensionless); requirements of normalizability of our
basis states constrain the parameters to the ranges$$0<\mu<\infty\ ,\quad
-1<2\,\beta<\infty\ .$$Therein, the orthonormality of the generalized
Laguerre polynomials, inherent to their definition, is equivalent to the
orthonormality of the radial basis functions $\Phi_{k,\ell}(r)$:
$$\int\limits_0^\infty{\rm d}r\,r^2\,\Phi_{k,\ell}(r)\,\Phi_{k',\ell}(r)
=\delta_{kk'}\ ,\quad k,k'=0,1,\dots\ ;$$this condition fixes the
normalization constant $N_{k,\ell}^{(\mu,\beta)}$~to
$$N_{k,\ell}^{(\mu,\beta)}=\sqrt{\frac{(2\,\mu)^{2\,\ell+2\,\beta+1}\,k!}
{\Gamma(2\,\ell+2\,\beta+k+1)}}\ .$$

\newpage

Fortunately the assumed factorization of every basis function persists in its
momentum-space representation:$$\widetilde\psi_{k,\ell
m}(\mbox{\boldmath{$p$}})=\widetilde\Phi_{k,\ell}(p)\,{\cal Y}_{\ell
m}(\Omega_p)\ ,\quad p\equiv|\mbox{\boldmath{$p$}}|\ .$$Analytical statements
about Hamiltonians that involve a kinetic-energy operator nonlocal in
configuration space, such as a relativistic square root (\ref{Eq:RKE}), are
facilitated by an explicit knowledge of the momentum-space basis states. One
of the great advantages of the generalized-Laguerre basis is the availability
of its {\em analytic\/} Fourier transform.

For all factorizations into radial and angular parts, as consequence of the
Fourier transformation acting on the Hilbert space $L_2(R^3)$ of the
square-integrable functions on the three-dimensional space $R^3,$ the radial
parts of all basis functions that represent the chosen basis vectors in
configuration space and momentum space, respectively, are related by
so-called Fourier--Bessel
transformations:\begin{eqnarray*}\Phi_{k,\ell}(r)&=&{\rm
i}^\ell\,\sqrt{\frac{2}{\pi}}\int\limits_0^\infty{\rm
d}p\,p^2\,j_\ell(p\,r)\,\widetilde\Phi_{k,\ell}(p)\
,\\\widetilde\Phi_{k,\ell}(p)&=&(-{\rm
i})^\ell\,\sqrt{\frac{2}{\pi}}\int\limits_0^\infty{\rm
d}r\,r^2\,j_\ell(p\,r)\,\Phi_{k,\ell}(r)\ ,\\&&\quad\mbox{for all}\
k=0,1,\dots,\ \ell=0,1,\dots\ ;\end{eqnarray*}the angular-integration
remnants $j_n(z)$ ($n=0,\pm 1,\dots$) label the spherical Bessel functions of
the first kind \cite{Abramowitz}. For the generalized-Laguerre basis under
consideration, these radial basis functions become in momentum space
\begin{eqnarray*}\widetilde\Phi_{k,\ell}(p)&=&
N_{k,\ell}^{(\mu,\beta)}\,\frac{(-{\rm i})^\ell\,p^\ell}
{2^{\ell+1/2}\,\Gamma\left(\ell+\frac{3}{2}\right)}\\
&\times&\sum_{t=0}^k\,\frac{(-1)^t}{t!}
\left(\begin{array}{c}k+2\,\ell+2\,\beta\\k-t\end{array}\right)\\
&\times&\frac{\Gamma(a_{t,\ell;\beta})\,(2\,\mu)^t}
{(p^2+\mu^2)^{a_{t,\ell;\beta}/2}}\\
&\times&F\left(\frac{a_{t,\ell;\beta}}{2},-\frac{\beta+t}{2};\ell+\frac{3}{2};
\frac{p^2}{p^2+\mu^2}\right),\end{eqnarray*}with the hypergeometric series
$F(u,v;w;z)$, defined,~in terms of the gamma function $\Gamma,$ by the power
series \cite{Abramowitz}\begin{eqnarray*}&&F(u,v;w;z)\\
&&=\frac{\Gamma(w)}{\Gamma(u)\,\Gamma(v)}\,\sum_{n=0}^\infty\,
\frac{\Gamma(u+n)\,\Gamma(v+n)}{\Gamma(w+n)}\,\frac{z^n}{n!}\
,\end{eqnarray*}and the simplifying abbreviation
$a_{t,\ell;\beta}\equiv2\,\ell+\beta+t+2.$

Clearly, the momentum-space radial basis functions
$\widetilde\Phi_{k,\ell}(p)$ have to satisfy the orthonormalization
condition$$\int\limits_0^\infty{\rm
d}p\,p^2\,\widetilde\Phi_{k,\ell}^\ast(p)\,\widetilde\Phi_{k',\ell}(p)
=\delta_{kk'}\ ,\quad k,k'=0,1,\dots\ .$$


\begin{thebibliography}{30}
\bibitem{BSE}E.~E.~Salpeter and H.~A.~Bethe, Phys.~Rev.~{\bf 84} (1951) 1232.
\bibitem{SE}E.~E.~Salpeter, Phys.~Rev.~{\bf 87} (1952) 328.
\bibitem{Lucha91:BSQ}W.~Lucha, F.~F.~Sch\"oberl, and D.~Gromes, Phys.\
Rep.~{\bf 200} (1991) 127.
\bibitem{Resag94}J.~Resag {\it et al.}, Nucl.~Phys.~A {\bf 578} (1994) 397
[nucl-th/9307026].
\bibitem{Kopaleishvili01}T.~Kopaleishvili, Phys.~Part.~Nucl.\ {\bf 32} (2001)
560 [hep-ph/0101271].
\bibitem{Lucha:RVT}W.~Lucha and F.~F.~Sch\"oberl, Phys.~Rev.~Lett.~{\bf 64}
(1990) 2733.
\bibitem{Lucha:RVTs}W.~Lucha and F.~F.~Sch\"oberl, Mod.~Phys.~Lett.~A~{\bf 5}
(1990) 2473.
\bibitem{Fock30}V.~Fock, Z.~Phys.~{\bf 63} (1930) 855.
\bibitem{Brack83}M.~Brack, Phys.~Rev.~D {\bf 27} (1983) 1950.
\bibitem{Lucha94:Como}W.~Lucha and F.~F.~Sch\"oberl, in: Proc.~Int.~Conf.\ on
{\it Quark Confinement and the Hadron Spectrum}, edited by N.~Brambilla and
G.~M.~Prosperi (World Scientific, River Edge (N.~J.), 1995) p.~100
[hep-ph/9410221].
\bibitem{Lucha98:Dubna}W.~Lucha and F.~F.~Sch\"oberl, in: Proc.~XI$^{\rm th}$
Int.\ Conf.~{\it Problems of Quantum Field Theory}, editors: B.~M.~Barbashov,
G.~V.~Efimov, and A.~V.~Efremov (Joint Institute f.~Nuclear Research, Dubna,
1999) p.~482 [hep-ph/9807342].
\bibitem{Lucha:Oberwoelz}W.~Lucha and F.~F.~Sch\"oberl, Int.~J.~Mod.~Phys.~A
{\bf 14} (1999) 2309 [hep-ph/9812368].
\bibitem{Lucha:Dubrovnik}W.~Lucha and F.~F.~Sch\"oberl, Fizika B {\bf 8}
(1999) 193 [hep-ph/9812526].
\bibitem{Reed78}M.~Reed and B.~Simon, {\em Methods of Modern Mathematical
Physics~IV: Analysis~of Operators\/} (Academic Press, New York, 1978)
Section~XIII.1.
\bibitem{Weinstein72}A.~Weinstein and W.~Stenger, {\em Methods of
Intermediate Problems for Eigenvalues -- Theory and Ramifications\/}
(Academic Press, New York, 1972) Chapters 1 and 2.
\bibitem{Thirring90}W.~Thirring, {\em A Course in Mathematical Physics~3:
Quantum Mechanics of Atoms and Molecules\/} (Springer, New York/Wien, 1990)
Section~3.5.
\bibitem{Lucha96rcpaubel}W.~Lucha and F.~F.~Sch\"oberl, Phys.~Rev.~A {\bf 54}
(1996) 3790 [hep-ph/9603429].
\bibitem{Lucha99-1dimsrcp}W.~Lucha and F.~F.~Sch\"oberl, J.~Math.~Phys.~{\bf
41} (2000) 1778 [hep-ph/9905556].
\bibitem{Martin88}A.~Martin, Phys.~Lett.~B {\bf 214} (1988) 561.
\bibitem{Lucha00-HO}R.~L.~Hall, W.~Lucha, and F.~F.~Sch\"oberl, J.~Phys.~A
{\bf 34} (2001) 5059 [hep-th/0012127].
\bibitem{Lucha01-DMAIa}R.~L.~Hall, W.~Lucha, and F.~F.~Sch\"oberl, J.~Math.\
Phys.~{\bf 42} (2001) 5228 [hep-th/0101223].
\bibitem{Lucha01-DMAIb}R.~L.~Hall, W.~Lucha, and F.~F.~Sch\"oberl, Int.~J.\
Mod.~Phys.~A {\bf 17} (2002) 1931 [hep-th/0110220].
\bibitem{Lucha02-DMAII}R.~L.~Hall, W.~Lucha, and F.~F.~Sch\"oberl, Int.~J.\
Mod.~Phys.~A {\bf 18} (2003) 2657 [hep-th/0210149].
\bibitem{Lucha02-sum}R.~L.~Hall, W.~Lucha, and F.~F.~Sch\"oberl, J.~Math.\
Phys.~{\bf 43} (2002) 5913 [math-ph/0208042].
\bibitem{Lucha02-CP}R.~L.~Hall, W.~Lucha, and F.~F.~Sch\"oberl, in: Proc.\
Int.~Conf.~on {\it Quark Confinement and the Hadron Spectrum V},
eds.~N.~Brambilla and G.~M.~Prosperi (World Scientific, Singapore, 2003)
p.~500.
\bibitem{Herbst}I.~W.~Herbst, Commun.~Math.~Phys.~{\bf 53} (1977) 285; {\it
ibid}.~{\bf 55} (1977) 316 (addendum).
\bibitem{Martin89}A.~Martin and S.~M.~Roy, Phys.~Lett.~B {\bf 233} (1989)
407.
\bibitem{Lucha01-NHO}R.~L.~Hall, W.~Lucha, and F.~F.~Sch\"oberl, J.~Math.\
Phys.~{\bf 43} (2002) 1237; {\it ibid}.~{\bf 44} (2003) 2724 (E)
[math-ph/0110015].
\bibitem{Lucha03-NHO-m=0}R.~L.~Hall, W.~Lucha, and F.~F.~Sch\"oberl, Phys.\
Lett.~A {\bf 320} (2003) 127 [math-ph/0311032].
\bibitem{Lucha04-NV(r2)}R.~L.~Hall, W.~Lucha, and F.~F.~Sch\"oberl, J.~Math.\
Phys.~{\bf 45} (2004) 3086 [math-ph/0405025].
\bibitem{Jacobs86}S.~Jacobs, M.~G.~Olsson, and C.~Suchyta III, Phys.\ Rev.~D
{\bf 33} (1986) 3338; {\it ibid}.~{\bf 34} (1986) 3536 (E).
\bibitem{Lucha:LagB}W.~Lucha and F.~F.~Sch\"oberl, Phys.~Rev.~A {\bf 56}
(1997) 139 [hep-ph/9609322].
\bibitem{Lucha94:VA-SRCP}W.~Lucha and F.~F.~Sch\"oberl, Phys.~Rev.~D {\bf 50}
(1994) 5443 [hep-ph/9406312].
\bibitem{Lucha96:CCC}W.~Lucha and F.~F.~Sch\"oberl, Phys.~Lett.~B {\bf 387}
(1996) 573 [hep-ph/9607249].
\bibitem{Lucha92:QAQBS}W.~Lucha and F.~F.~Sch\"oberl,
Int.~J.~Mod.~Phys.~A~{\bf 7} (1992) 6431.
\bibitem{Lucha:Q/A}W.~Lucha and F.~F.~Sch\"oberl, Phys.~Rev.~A {\bf 60}
(1999) 5091 [hep-ph/9904391].
\bibitem{Lucha:A/Q}W.~Lucha and F.~F.~Sch\"oberl, Int.~J.~Mod.~Phys.~A {\bf
15} (2000) 3221 [hep-ph/9909451].

\newpage

\bibitem{Lucha92:MAP}W.~Lucha, H.~Rupprecht, and F.~F.~Sch\"oberl, Phys.\
Rev.~D {\bf 45} (1992) 1233.
\bibitem{Lucha:SAMM}W.~Lucha and F.~F.~Sch\"oberl, Int.~J.~Mod.~Phys.~C {\bf
11} (2000) 485 [hep-ph/0002139].
\bibitem{Abramowitz}{\it Handbook of Mathematical Functions}, eds.~M.\
Abramowitz and I.~A.~Stegun (Dover, New York, 1964).
\bibitem{Bateman}Bateman Manuscript Project, A.~Erd\'elyi {\it et al}., {\em
Higher Transcendental Functions} (McGraw--Hill, New York, 1953) Volume~II.
\end{thebibliography}
\end{document}